\documentclass[prl,aps,twocolumn,nofootinbib,floatfix]{revtex4-2}
\usepackage{graphicx}
\usepackage{bm}
\usepackage{hyperref}

\newcommand{\cov}[1]{C^{(#1)}}
\def\be{\begin{equation}}
\def\ee{\end{equation}}
\def\ba{\begin{eqnarray}}
\def\ea{\end{eqnarray}}
\def\half{\frac{1}{2}}

\def\part{\partial}

\def\a{\alpha}
\def\b{\beta}
\def\gam{\gamma}
\def\Gam{\Gamma}

\def\ome{\omega}

\def\eps{\epsilon}

\def\la{\langle}
\def\ra{\rangle}

\def\bin{b_{in}}
\def\bout{b_{out}}

\def\tbout{\tilde{b}_{out}}
\def\ta{\tilde{a}}

\def\Dt{T}

\def\kmax{k_{\max}}
\begin{document}

\title{
Entanglement production in the decay of a metastable state}
\author{Sergei Khlebnikov}
\affiliation{Department of Physics and Astronomy, Purdue University, West Lafayette, 
IN 47907, USA}

\begin{abstract}
When a metastable state decays into radiation, there must be entanglement between 
the radiation and the decaying system, as well as between radiation collected at late 
and early times. We study the interplay between these two types of entanglement
in simple Gaussian models in the Markov approximation.
We define, via a windowed Fourier transform, multimode quantum states associated with
radiation fragments produced at different times and compute the 
corresponding entanglement entropy increments. On the basis of these results, we argue
that such entropy increments are useful entanglement measures, especially 
in cases, such as Hawking
radiation, where one wishes to separate the radiation into ``old'' and ``new.''
\end{abstract}

\maketitle

Decay of a metastable state in quantum mechanics is a probabilistic process and, as such,
must have an amount of entropy associated with it. This entropy reflects the uncertainty 
as to how much of the state has already decayed by some time $t$. 
If, for example,
the decaying system (subsystem $A$) starts in a pure quantum state and ends in a pure state as 
well, a properly defined entropy, $S_A(t)$, should be zero at $t =0$ and
$t\to \infty$, where we know the state vector with certainty, but nonzero at intermediate
times. 
A natural candidate for such a definition is the entropy of entanglement
of the metastable system with the decay products.

A common reason why quantum systems decay is that they interact with radiation; by 
radiation
we mean here a subsystem with a continuous (or quasicontinuous) energy spectrum. 
The simplest way in which
such a decay can occur is by free streaming, as when, for instance, quanta populating 
a single mode of an optical resonator leak out through one of the end mirrors.
The Hamiltonian describing this system is
\be
H = \ome_s a^\dagger a + i \sum_\nu g_\nu ( b_\nu^\dagger a - a^\dagger b_\nu) 
+ \sum_\nu \eps_\nu b_\nu^\dagger b_\nu \, ,
\label{ham}
\ee
where the operators $a$ and $a^\dagger$ correspond to the resonator, and $b_\nu$, $b_\nu^\dagger$
to the outside radiation. The parameters $\ome_s$ and $g_\nu$ are assumed real and positive.
We set $\hbar = 1$ everywhere. Much of our discussion will focus on this system, although
towards the end we will consider also the case when some other system decays into
$a$ and $b$ by parametric resonance; this corresponds to replacing 
$b_\nu^\dagger a$ in (\ref{ham}) with $b_\nu^\dagger a^\dagger$. 

When the frequency $\ome_s$ of the resonator lies high in the radiation spectrum,
while the decay width due to the loss of quanta is small, there is often an
additional simplification: one may be able to replace
the radiation spectral density (including powers of the
coefficients $g_\nu$) with a constant. In the time domain, this corresponds to
\be
\sum_\nu g_\nu^2 e^{-i \eps_\nu (t-t')} \to \Gam \delta(t - t') \, ,
\label{mark}
\ee
where $\Gam$ is a positive constant. This is known as the Markov approximation.
We will use it, as well as the choice of Gaussian initial states, as our
simplifying assumptions.

We assume throughout that radiation as a whole starts in the 
vacuum of $b_\nu$, while the resonator
starts in a squeezed vacuum or in a thermal state. 
Computation of the entanglement entropy of the resonator itself then
presents no problem:
it amounts to applying a Bogoliubov transformation to the
$2\times 2$ covariance matrix of $A$ (equal here to the symmetric part of the equal-time 
correlator $\la X_\a X_\b \ra$, where $X_\a$, $\a = 1,2$, are a pair of independent
canonical variables).

Things are different for the radiation subsystem. Because
the spatial volume occupied by radiation is vast, one may prefer in this case to focus, 
instead of the total entropy, on the amount
deposited into radiation during a particular time interval. We refer to such an amount as an
{\em entropy increment}. 

While the total number of Fourier modes required to represent
radiation produced during a time interval is infinite, the modes are discrete,
and we can construct a
covariance matrix describing all these modes up to a certain cutoff. The entropy of the
resulting multimode Gaussian state can then be computed by symplectic diagonalization 
\cite{Williamson,Simon&al} of that matrix. We may expect the precise
value of the (sufficiently large) cutoff to be unimportant, provided most
of the entropy is deposited into a finite bandwidth.

The main result of this paper is that focusing on entropy increments, rather 
than on conventionally defined entropies, provides
a good way of thinking about entanglement entropy of radiation, in the sense that (i) it
is consistent with intuition in simpler cases, and (ii) it can augment that intuition in 
cases where one looks for a finer characterization of the
entanglement, as for instance when one wishes to treat separately amounts of 
radiation collected during different periods of time.

In general, outgoing radiation wavepackets can be constructed from the interaction
picture operators that coincide with the Heisenberg operators $b_\nu(t_+)$ at some future
moment $t_+$, as follows:
\be
\bout(t) = \sum_\nu h_\nu e^{-i \eps_\nu (t - t_+)} b_\nu(t_+) \, ,
\label{bout_def}
\ee
where $h_\nu$ are coefficients. In the Markov approximation, 
there is a particularly useful choice
$h_\nu = g_\nu / \sqrt{\Gam}$, for which $\bout(t)$ are simply related to the input operators
and obey the commutation 
relation $[\bout(t), \bout^\dagger(t')] = \delta(t-t')$ 
\cite{Gardiner&Collett}. As a result, in this case, we can define mutually commuting 
radiation subsystems via Fourier transforms with sharply delineated time 
windows [see Eq.~(\ref{Bk}) below]. We expect that separation of radiation into
temporal subsystems is possible also outside the Markov approximation, although 
then one may 
have to use a different set of orthogonal functions (e.g., wavelets), and the resulting 
subsystems may not have as sharply defined boundaries.

In what follows, we refer to the radiation collected during the interval $(0,t_0)$ 
as ``old'' and to that collected during $(t_0, t)$ as ``new.'' These will be our
subsystems $B_1$ and $B_2$, respectively. We also compute the entropies ($S$) of the 
compound subsystems $B_1 B_2$ and $B_2 A$, where $A$ is the resonator 
at time $t$ (the time we finish collecting new radiation).
Note that, while $S_{B_1B_2}$ is computed using two separate Fourier 
transforms, one for $B_1$ and the other for $B_2$, it takes into account 
entanglement between $B_1$ and $B_2$. This can be viewed as entanglement 
``in time'' across the boundary at $t = t_0$. Our notation $S_{B_1B_2}(0,t_0,t)$ for that
increment includes all three time labels.

Note that the subsystems $B_1$ and $B_2$ are composed of radiation
modes emitted during finite time
intervals, so the entanglement-in-time as measured by the entropy increments is distinct 
from the recently proposed measures characterizing entanglement of subsystems 
located sharply at two different moments 
of time (see Ref.~\cite{Milekhin&al} and references therein). 
For example, the increments computed here are real-valued, while those other measures 
are in general complex.

We focus on
the following relations, all of which have been obtained numerically.
 
(i) {\em Conservation of uncertainty}. Regardless of whether the initial and final
states of the resonator ($A$) are pure or mixed, we find
\be
S_{B_2 A}(t_0, t) = S_A(t_0) \, ,
\label{cons}
\ee
meaning in particular that the left-hand side is independent of $t$. In other words,
production of new radiation ($B_2$)
does not change our uncertainty about the state of the 
compound $B_2A$ system.

(ii) Regardless of whether the initial and final states of $A$ are pure, 
$S_{B_1B_2}(0, t_0, t)$ is independent of $t_0$. 

(iii) If the initial state of $A$ is pure, then in addition to (ii) we have
\be
S_{B_1B_2}(0, t_0, t) = S_A(t) 
\label{ii}
\ee
for any $t > t_0$. Eq.~(\ref{ii}) is consistent with thinking of $B_1B_2$ and $A$ as 
two subsystems whose union is in a pure state. Indeed, we also compute $S_{B_1B_2A}$ and 
find that, in this case, the results are
consistent with the condition of overall purity,
$S_{B_1B_2A} = 0$ at all times.

(iv) More generally, $S_{B_1B_2A} = S_A(0)$. 
If the final state of $A$ is pure, this becomes
\be
\lim_{t\to\infty} S_{B_1 B_2} (0, t_0, t) = S_A(0) \, ,
\label{ff}
\ee
meaning that any uncertainty 
we had in the state of $A$ at $t=0$ must be eventually absorbed by the radiation.

Let us note that the possibility of assigning well-defined
quantum states to individual radiation 
fragments is not obvious {\em a priori}. Recall that
the symplectic eigenvalues
$s_\ell$ of a covariance matrix corresponding to a
quantum state must satisfy the condition $s_\ell \geq 1$ 
(in the present normalization), 
following from the uncertainty principle \cite{Simon&al}. Here, we find that the
covariance matrices of the 
subsystems $B_2 A$ and $B_1 B_2$ do satisfy this criterion, within expected 
numerical accuracy. Moreover, we find that only one symplectic eigenvalue of each of
these matrices is distinct from unity, which means that, in the relations above,
not only the entropies match, but all the nonzero elements of the density matrices also
do, although our evidence for that at this point is purely numerical.
It would certainly be interesting to have an analytical confirmation.

The existence of
quantum states associated with radiation fragments suggests
that various inequalities, characteristic of the entanglement entropy, will 
apply here as well.
This concerns, in particular, the strong subadditivity inequality proven in 
Ref.~\cite{Lieb&Ruskai} (for related inequalities, see Ref.~\cite{Carlen&Lieb}).
While a detailed study of the question is beyond the scope of this paper, in the few cases 
where we checked, strong subadditivity in the form
\be
S_{B_1 B_2} + S_{B_2 A} \geq S_{B_2} + S_{B_1 B_2 A} 
\label{subadd}
\ee
was satisfied.

Separation of radiation into old and new is at the heart of a famous
paradox---the apparent loss of information during the decay of a black hole into
Hawking radiation \cite{Hawking:1976}; for a recent review, 
see Ref.~\cite{Almheiri&al:review}. 
In particular, the differential
version of our Eq.~(\ref{cons}), $d S_{B_2 A}/ dt = 0$, is similar to the 
``no drama'' condition employed in some discussions of the paradox
\cite{Mathur,Almheiri&al:firewall}. The analogy is particularly close for
the case when pairs of $a$ and $b$ quanta are produced by parametric resonance.
However, while we find that we can define separate subsystems for portions of 
the outgoing radiation ($B$) emitted at different times,
we see no way to similarly partition the ``interior'' 
$A$ quanta, as they are all deposited into the same mode.
In other words, we are not able to cut out a small fragment $A'$ of the interior 
radiation, so as to claim that it is maximally entangled with a fragment $B'$
produced concurrently on the outside,
$S_{A'B'} = 0$. A relation such as this, however, is crucial to formulation of 
the information paradox 
\cite{Mathur,Almheiri&al:firewall} on the basis of the strong subadditivity 
inequality of Ref.~\cite{Lieb&Ruskai}. In that context, it amounts to the assumption 
that there is an elementary pair-producing process that uses each time a new 
interior radiation mode ($A'$), initially in its vacuum state. 
In general, such an assumption will not hold  
if the total number of available modes is 
finite, and the process needs to reuse them. By the time $t_0$ a mode $A$
is reused, it has already been entangled with early radiation (or some other 
subsystem, if any). This implies $S_A(t_0) > 0$, just as we find here.

We now describe computations that lead to the conclusions presented above.
Application of the input-output theory \cite{Gardiner&Collett}
and the Markov approximation to (\ref{ham}) results
in the following equation of motion for $a(t)$:
\be
\dot{a} = -i \ome_s a - \half \Gam a  - \sqrt{\Gam} \bin
\label{eqm}
\ee
where 
\be
\bin(t) =  \frac{1}{\sqrt{\Gam}} \sum_\nu g_\nu e^{-i\eps_\nu t} b_\nu(0)
\label{bin}
\ee
is the input operator, representing radiation just before it interacts with the resonator
at time $t$,
while (\ref{bout_def}) with $h_\nu = g_\nu / \sqrt{\Gam}$
is the output operator, representing 
radiation just after that time. These satisfy 
the relation $\bout(t) = \sqrt{\Gam} a(t) + \bin(t)$, as well as
the commutation relations $[\bin(t), \bin^\dagger(t')] = \delta(t -t')$ and 
a similar one for $\bout$.

We consider only cases when the radiation is initially in the trivial vacuum annihilated 
by $\bin$. 
Upon solution of the linear Eq.~(\ref{eqm}),
the state of the resonator is conveniently represented by the covariance matrix 
\be
\cov{XX}_{\a\b}(t) = \half \la \{ X_\a(t) , X_\b(t) \} \ra 
\label{cov_def}
\ee
of the Hermitian quadratures  
\ba
X_1(t) & = &   \ta(t) +   \ta^\dagger(t)   \, , \label{X1} \\
X_2(t) & = &  i [  \ta^\dagger(t) -  \ta(t) ] \, , \label{X2}
\ea
where $\ta(t)$ are the rotating frame operators $\ta(t) = e^{i\ome_s t} a(t)$, and the
braces in (\ref{cov_def}) denote an anticommutator. The result is
\be
\cov{XX}_{\a\b}(t) = \left[ \cov{XX}_{\a\b}(0) - \delta_{\a\b} \right] e^{- \Gam t} + \delta_{\a\b} 
\label{covX}
\ee
($\a,\b = 1,2$).
The entanglement entropy of the resonator
is then obtained from the symplectic eigenvalue $s(t) = [\det \cov{XX}_{\a\b}(t)]^{1/2}$ 
of (\ref{covX}) by means of the ideal-gas formula
\be
S_A(t) = (n + 1) \ln (n + 1)  - n \ln n \, ,
\label{SA}
\ee
where $n = \half [s(t) - 1]$.
 
The companion calculation for the output quadratures
\ba
Z_1(t) & = &   \tbout(t) +   \tbout^\dagger(t)   \, , \label{Z1} \\
Z_2(t) & = &  i [  \tbout^\dagger(t) -  \tbout(t) ] \, ,  \label{Z2}
\ea
where $\tbout(t) = e^{i\ome_s t} \bout(t)$,
requires that we first discretize the signal on a time interval $(t_1, t_2)$. 
Following Ref.~\cite{paramp}, we use the windowed cosine transform
\be
Z_{\a k} = \frac{\eta_k}{\sqrt{\Dt}}
\int_{t_1}^{t_2} Z_\a(t) \cos \left[ \ome_k (t - t_1) \right]  dt \, ,
\label{Bk}
\ee
where $\Dt = t_2 - t_1$, 
$\eta_0 = 1$, $\eta_k = \sqrt{2}$ for $k > 0$, and
\be
\ome_k = \pi k / \Dt \, , \hspace{3em} k= 0,\dots, \kmax \, .
\label{ome}
\ee
The result for the covariance matrix of $Z_{\a k}$ can then be written as
\be
\cov{ZZ}_{\a\b,kk'} = \left[ \cov{XX}_{\a\b}(0) - \delta_{\a\b} \right] f_k f_{k'} 
+ \delta_{\a\b} \delta_{kk'} \, ,
\label{covZ}
\ee
where
\be
f_k = \sqrt{\frac{\Gam}{\Dt}} \frac{\gam e^{-\gam t_1} \eta_k}{\gam^2 + \ome_k^2} 
(1 - e^{-\gam \Dt} \cos \ome_k \Dt) \, ,
\label{fk}
\ee
and $\gam \equiv \Gam / 2$. Note the similarity between (\ref{covZ}) and (\ref{covX}).

Similar expressions are obtained for the cross covariances between $Z$ and $X$,
as well as those between the discretized signals corresponding to radiation collected
over different time intervals. They play a role in the computations for compound
systems, such as $B_2 A$ and $B_1 B_2$ discussed earlier.

The multimode
covariance matrix so obtained 
is brought to the Williamson normal form \cite{Williamson,Simon&al} numerically by a 
symplectic transformation, and the associated entropy increment is obtained by summing 
up all expressions of the form (\ref{SA}) with $n = \half (s - 1)$, where $s$ are the
symplectic eigenvalues.  

\begin{figure}
\includegraphics[width=3.25in]{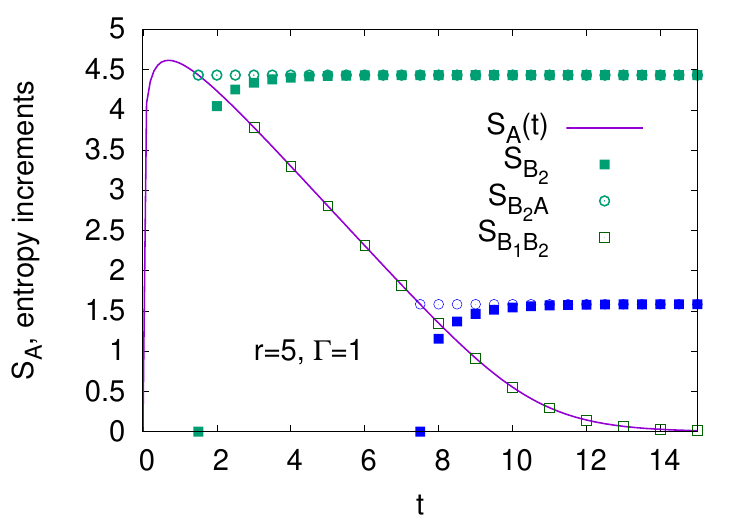}
\caption{\small Radiation entropy increments as functions of time 
for the case when
the resonator ($A$) starts in the squeezed vacuum corresponding to 
(\ref{init}) with $r=5$, while radiation ($B$) starts in 
the trivial vacuum. The solid line is 
the entanglement entropy of the 
resonator, $S_A(t)$. Eq.~(\ref{cons}) is verified here for two values of
$t_0$ ($t_0 = 1.5$ and $7.5$), with $S_{B_2A}(t_0,t)$ 
represented in both cases by empty circles.
$S_{B_1 B_2}(0,t_0,t)$ (empty squares) is found to be independent of $t_0$, and 
is shown here for $t_0 =3$. As a function of $t$, it follows $S_A(t)$, 
in accordance with (\ref{ii}).
}
\label{fig:lim}
\end{figure}

In Fig.~\ref{fig:lim}, we plot the entanglement entropy $S_A$ of the resonator and 
various entropy increments as functions of the time $t$ at which we stop collecting
radiation, for the case when the resonator starts out in the squeezed vacuum with 
the covariance matrix
\be
\cov{XX}_{\a\b}(0) = \mbox{diag} (e^{2r}, e^{-2r}) \, .
\label{init}
\ee
The computed entropy increments correspond to the  
wavenumber cutoff $\kmax = 200$; increasing $\kmax$ further does not produce 
visible changes in the plots. Since both the initial and final states of
$A$ in this case are pure, we expect all three relations (\ref{cons})--(\ref{ff})
to apply, and we see that they do. 

\begin{figure}
\includegraphics[width=3.25in]{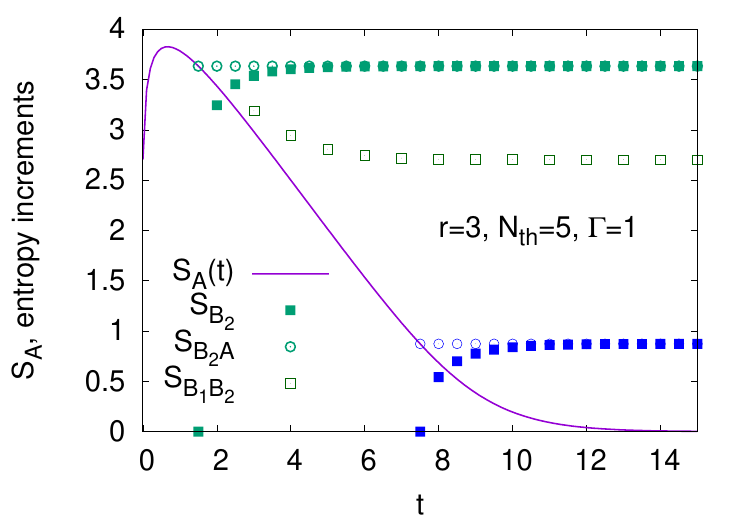}
\caption{\small Same as in Fig.~\ref{fig:lim}, but for a case when the initial
state of $A$ is mixed (a squeezed thermal state with population $N_{th}$). 
$S_{B_1 B_2}(0,t_0,t)$ (empty squares) no longer follows
$S_A(t)$ (solid line) but still approaches $S_A(0)$ at large $t$, 
in accordance with (\ref{ff}).}
\label{fig:thermal}
\end{figure}

In Fig.~\ref{fig:thermal}, we show the same set of quantities for the case when the
initial state of the resonator is a squeezed thermal state, with the covariance 
matrix obtained by multiplying (\ref{init}) with 
$2 N_{th} + 1$, where $N_{th}$ is the initial thermal population. The main difference
with Fig.~\ref{fig:lim} is that the relation (\ref{ii}) is no longer satisfied,
as indeed it is not expected to. However, (\ref{ff}) is now nontrivial,
and it is confirmed by the plot.

The approach of $S_{B_2}(t_0,t)$ to saturation at large times is well fit by 
a $\mbox{const}\times e^{-\Gam t}$, with the exponent independent of both $t_0$ 
and the initial state. This is not obvious from (\ref{fk}), as
the second term in the bracket there decays with $T = t - t_0$ with only half 
the rate $\Gam$. The speedup to $e^{-\Gam t}$ is then apparently a result of
correlations between 
different Fourier modes (whose frequency spacing is also controlled by $T$).

Note that in both Figs.~\ref{fig:lim} and
\ref{fig:thermal}, $S_{B_2}$ approaches $S_{B_2 A}$ at $t\to\infty$. This is 
a consequence of the resonator approaching a pure state (vacuum) at 
large times. We may also
consider a case when that is not so. Consider the Hamiltonian
\be
H' = -\ome_s a^\dagger a + i \sum_\nu g_\nu (b_\nu^\dagger a^\dagger  - a b_\nu) 
+ \sum_\nu \eps_\nu b_\nu^\dagger b_\nu \, ,
\label{ham2}
\ee
obtained from (\ref{ham}) by replacing $b_\nu^\dagger a$ with $b_\nu^\dagger a^\dagger$
and reversing the sign of the first term ($\ome_s > 0$). 
This describes decay of some other system (a ``condensate'') into pairs of
$a$ and $b$ quanta
by parametric resonance, in the regime where the backreaction has not yet set in,
so the amplitude of the condensate is constant and can be absorbed into the parameters
$g_\nu$. Applying the input-output theory \cite{Gardiner&Collett}
and the Markov approximation, we have, in place of (\ref{eqm}), 
\be 
\dot{a} = i \ome_s a + \half \Gam a + \sqrt{\Gam} \bin^\dagger \, ,
\label{eqm2}
\ee
where $\Gam > 0$ and $\bin$ are given by the same expressions as before, while 
the output operators now are
\be
\bout = \bin + \sqrt{\Gam} a^\dagger \, .
\label{bout2}
\ee 
Note that
the term proportional to $\Gam$ in (\ref{eqm2}) now corresponds to amplification,
and it is $a^\dagger$ rather than $a$ that appears in (\ref{bout2}). 
The covariance matrices have to be recalculated accordingly.

\begin{figure}
\includegraphics[width=3.25in]{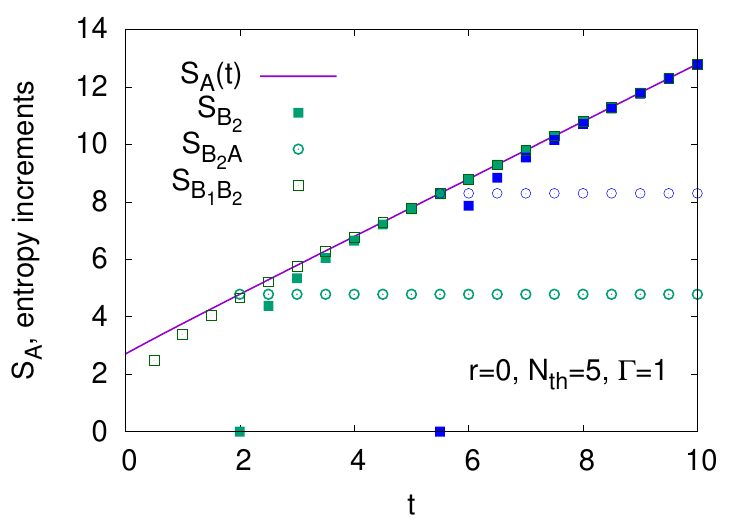}
\caption{\small Conservation of uncertainty, Eq.~(\ref{cons}), for the case when
the resonator ($A$) starts in a thermal state, and radiation ($B$) starts in 
vacuum, and both are amplified by 
parametric resonance. As before, empty circles represent
$S_{B_2 A}(t_0,t)$ for two values of $t_0$ (here, $t_0 = 2$ and $5.5$). 
The plots verify the relation (\ref{cons}) for a case
when neither the initial nor the final state of the resonator is pure.}
\label{fig:amp}
\end{figure}

Fig.~\ref{fig:amp} is the counterpart of Fig.~\ref{fig:lim} for the case when 
the Hamiltonian (\ref{ham2}) amplifies a thermal initial state of $A$. 
We see that Eq.~(\ref{cons}) still 
holds. Note that, in contrast to the previous case (decay to vacuum), now 
$S_{B_1 B_2} \geq S_{B_2}$ for all $t$ and $t_0$. Identifying $S_{B_1 B_2 A}$ with
$S_A(0)$ (as confirmed numerically) and observing that $S_{B_2 A}$ in Fig.~\ref{fig:amp}
lies above $S_A(0)$, we see
how the strong subadditivity condition (\ref{subadd}) can 
be maintained in the presence of pair production.

To summarize, we have considered decays of simple physical systems, using
a windowed Fourier transform to compute the entanglement
entropy (EE) increments associated with radiation emitted during finite time
intervals. These increments
are different from the conventional EE, which is defined with respect to 
a single moment of time, but we find that they share some of its properties 
(such as $S_B = S_A$ for two subsystems whose union is in a pure state).
On the other hand, we expect the increments to provide a finer characterization of 
entanglement in 
cases, such as Hawking radiation, where one aims to treat separately the old
and new radiation fragments.

\end{document}